\documentstyle{article}
\newcommand{\lab}{\label}
\newcommand{\bc}{\begin{center}}
\newcommand{\ec}{\end{center}}
\newcommand{\noi}{\noindent}
\newcommand{\barr}{\begin{array}}
\newcommand{\bey}{\begin{eqnarray}}
\newcommand{\be}{\begin{equation}}
\newcommand{\ear}{\end{array}}
\newcommand{\eey}{\end{eqnarray}}
\newcommand{\ee}{\end{equation}}

\newcommand{\pde}{\partial}

\newcommand{\spao}[1]{\mbox{\hspace{#1}}}
\newcommand{\spav}[1]{\parbox{1mm}{\vspace*{#1}}}

\newcommand{\ssty}{\scriptstyle}
\newcommand{\sssty}{\scriptscriptstyle}

\newcommand{\text}{\textstyle} 

\newsavebox{\ipiu}
\newsavebox{\imen}

\renewcommand{\a}{\alpha}
\renewcommand{\b}{\beta}

\newcommand{\da}{\dagger}
\renewcommand{\d}{\delta}

\newcommand{\ov}{\over}
\newcommand{\ot}{\otimes}

\sbox{\ipiu}{$\ssty i \sssty +1$}
\sbox{\imen}{$\ssty i \sssty -1$}

\begin{document}


\rightline{DPS-NA-19/95}
\rightline{INFN-NA-IV-19/95}
\rightline{DFUPG-102/95}

\bc \spav{1cm}\\
{\LARGE\bf  Knizhnik-Zamolodchikov equation and extended symmetry for stable
Hall states }
\bigskip

{\large A. De Martino \\
{\normalsize\em GNSM Sez. di Perugia\\}
{\normalsize\em c/o Dipartimento di Fisica, Universit\`a di Perugia\\}

 and \\ R. Musto  \\}
{\normalsize\em Dipartimento di Scienze Fisiche, Universit\`a di Napoli\\}
{\normalsize\em and INFN Sez. di Napoli\\}
\spav{1.6cm}\\
{\small\bf Abstract\\}
\spav{2mm}\\
{\small\parbox{13cm}{\spao{4mm}

We describe a $n$ component abelian Hall fluid as a system of {\it composite
bosons} moving
in an average null field given by the external magnetic field and  by the
statistical flux
tubes located at the position of the particles. The collective vacuum state,
in which the bosons condense, is characterized by a Knizhnik-Zamolodchikov
differential equation relative to a $\hat {U}(1)^n$ Wess-Zumino model.
In the case of states belonging to  Jain's sequences the Knizhnik-Zamolodchikov
equation naturally leads to the presence of an $\hat{U}(1)\ot \hat{SU}(n)$
extended  algebra. Only the $\hat{U}(1)$ mode is charged while
the $\hat{SU}(n)$ modes  are neutral, in agreement with
recent results obtained in the study of the edge states.

}\\}
\spav{4mm}\\
\bigskip
\bigskip
\bigskip
\baselineskip=12pt

\bigskip
\bigskip
\ec
\hfill

\newpage


It is well known that a simple physical idea lies behind the different
theoretical descriptions of the fractional quantum Hall effect (FQHE),
namely that an Hall fluid should be described not in terms of the ordinary
electrons but of the quasiparticles obtained by binding to them an
appropriate number of vortices.
This picture, clearly present in the Laughlin wave function$^1$,
widely discussed and exploited by Wilczeck$^2$ and formulated in the
framework of a Ginsburg-Landau model by several authors$^3$, finds a precise
mathematical formulation in two dimensional Conformal Field Theory (2DCFT),
where the non trivial braiding of two quasiparticles is realized as the
exchange of the corresponding Vertex Operators$^4$,
and in the Chern-Simons
(CS) lagrangian approach$^5$, that  gives the possibility
of going beyond the mean
field approximation, by evaluating the fluctuations of the CS field$^6$.

The most appealing realization of this physical idea has
been put forward by Jain$^7$, who has suggested the possibility of looking at
the FQHE for the electrons at filling $\nu=n / (2np \pm 1)$ as a
manifestation of the integer effect for {\it composite fermions} obtained
by attaching to each electron an even number of flux units opposite to the
external magnetic field.
While the most direct experimental support for this point of view
derives from the observation
that  the most prominent Hall
plateaux are seen at the fillings $\nu=n/(2n\pm 1)$ (principal sequence)
and at $\nu=n/(4n\pm 1)$, further evidence  has been gathered from
the analysis of the
energy gaps for such fillings$^8 $  and  from the study$^9$  of
properties  of the state at $\nu = 1/2$, the accumulation point of the
principal sequence.

A very simple theoretical implementation  of Jain's approach has been obtained
by studying the motion of each quasiparticle  in the presence of the external
magnetic field and of infinitesimally thin statistical flux tubes located
at the position of the other quasiparticles$^{10}$. This analysis,
that can be generalized to an
arbitrary abelian Hall fluid, leads naturally to the Laughlin
ground state wave function (gswf) and  has the advantage of preserving in
the general case the algebraic structures present  in the case of filling
$\nu=1$, that is well  understood in terms of a single particle
description.
As a consequence, in this framework, an easy proof has been given$^{10,11}$
of   the important result$^{12,13}$ that the Laughlin gswf can be
characterized as a highest weight state of the
$W_{1+\infty}$ algebra of the area preserving non singular diffeomorfisms.

However in this paper,
instead of promoting the electrons to {\it
composite fermions}, we will turn them into {\it composite
bosons} moving in an average null field made of the external magnetic field
and the statistical flux tubes attached to the particles.
In this approach, a different characterization of the ground state of
a generic abelian Hall fluid can be obtained and typical
properties of Jain's states can be unveiled.
More specifically, for the simple case $\nu=1/m$ the collective vacuum,
in which the composite bosons condense, is characterized as a solution of
the Knizhnik-Zamolodchikov (KZ) equation$^{14}$
for the correlators of a $\hat {U}(1)$ Wess-Zumino
field, making contact with the analysis
of the Hall effect in terms of 2DCFT.
This result can be easily generalized to the generic
$n$ component abelian Hall fluid,
where the KZ equation is relative to $\hat {U}(1)^n$ 2DCFT correlators.
More interesting is the case of states belonging to Jain's sequences,
where  the KZ equation
exhibits the presence of an $\hat {U}(1) \otimes \hat {SU}(n)$ extended
algebra.
This approach confirms in a simple way results obtained in
the framework of representation theory of the
$W_{1+\infty}$ algebra$^{15}$ and by studying the dynamics of the edge
states in the presence of disorder$^{16}$.
It also gives a direct evidence that only the $\hat{U}(1)$ mode
is charged while the remaining are neutral.

Let us start our discussion by recalling that in the
$\nu=1$ case the analytic part of the Laughlin wave function is easily obtained
out the single
particle states as a Slater determinant
\be
\chi_{\nu =1} (z_1, z_2, \dots, z_{N})=\sum_P (-1)^P (b_1^{\da}  )^{n_{i_{1}}}
(b_2^{\da}  )^{n_{i_{2}}}  \dots (b^{\da}  _{N} )^{n_{i_{N}}}\chi_0 \, ,
\lab{van}
\ee \noi
where $n_{i_1}, n_{i_2} \dots  n_{i_{N}}$ is
a permutation, $P$, of the non negative integers smaller than
the electron number $N$, $ b^{\da}_i$ and   $b_i$
are creation and annihilation operators in the Fock-Bargmann representation:
\be
b^{\da}_i  =z_i\, , \,\,\,\,\,\,\,\,\, b_i= \pde_i \, ,
\lab{FB}
\ee
and the vacuum state is defined by
$b_i\chi_0 =0$.
Following Jain$^7$, one can describe the state at filling $\nu=1/(2p+1)
\equiv 1/m$ as a system, at the effective  filling $\nu_{eff}=1$, of {\it
composite fermions} obtained by  binding $-2p$ units of flux to each
electron. By studying the motion of each
composite fermion in the presence of the external magnetic field and
of the other particles$^{10}$, one sees that the operators  $ b^{\da}_i$ and
$b_i$
are to be modified as follows:
\be
b_i^{\da}    =z_i\, , \,\,\,\,\,\,
b_i= \pde_i -2p\pde_i \sum_{j\neq i} \ln(z_i-z_j) = \pde_i -2p\sum_{j\neq i}
{1 \over z_i-z_j}\, .
\lab{FBi}
\ee
This implies that the Fock space is built out of an intrinsically
collective vacuum $\chi_c$ defined by the condition $b_i \chi_c =0$, and
that the ground state   has the same structure than the one
given in eq. \ref{van}, namely:
\be
\chi _{1 \over m} (z_1, z_2, \dots, z_{N})=
\sum_P (-1)^P (b_1^{\da}   )^{n_{i_{1}}}
(b_2^{\da}   )^{n_{i_{2}}}  \dots (b_{N}^{\da}   )^{n_{i_{N}}}\chi_c =
\prod_{i<j} (z_i -z_j )^m\, .
\lab{VAN}
\ee \noi

More generally, one can consider the case of a generic $n$ component abelian
Hall fluid characterized by a symmetric integer valued matrix $K$
with odd diagonal elements$^5$. Then the generalized Laughlin gswf,
corresponding to the ground state of the composite fermions,
is given by
\be
\chi_K  (\{z_i^I\})= \prod_I \sum_{P_I}(-1)^{P_I} (b_1^{I\da})^{n^I_{i_1}}
(b_2^{I \da})^{n^I_{i_2}}  \dots (b_N^{I \da})^{n^I_{i_{N_I}}}\chi_c\, ,
\lab{VANH}\ee \noi
where  $i$ labels the electrons in a
given component, $I$ the different components and
$n^I_{i_1}, n^I_{i_2} \dots
n^I_{i_{N_I}}$ is
a permutation, $P_I$, of the sequence of non negative integers smaller than
the
number $N_I$ of electrons in the component $I$.
We have introduced the new operators
\be
(b_i^I)^{\da} =z_i^I\, , \,\,\,\,\,\, ~
b_i^I= \pde_i^I -\pde_i^I\,\, {\sum}^{'} H_{IJ}\,\, \ln(z_i^I-z_j^J) =\pde_i^I
 -{\sum}^{'} {H_{IJ}\over z_i^I-z_j^J}\, ,
\lab{FBI}
\ee \noi
where $H_{IJ} = K_{IJ}-\d_{IJ}$,
${\sum}^{'}$ means the sum on all values of $j,J$ such that $z_j^J\neq
z_i^I$
and the collective vacuum  $\chi_c$ is  defined by
$b_i^I \chi_c=0$.

As we have already mentioned, in this paper we follow a variation of this
approach, by describing the system as made out of {\it composite bosons} rather
than composite fermions.
For example, in the case of filling $\nu = 1/(2p+1) \equiv 1/m$,
we bind $-m$ units of flux to the electrons rather than  $-2p$. Studying,
along the same lines,
the motion of the composite bosons one is  lead to the introduction of
the new set of operators
\be
B_i^{\da}    =z_i\, , \,\,\, \,\,\,~~
B_i= \pde_i -m\,\pde_i \sum_{j\neq i} \ln(z_i-z_j) = \pde_i -m\sum_{j\neq i}
{1 \over z_i-z_j}\, .
\lab{FBBi}
\ee
In this case the composite bosons will condense in a collective
vacuum state, corresponding to the Laughlin gswf,
defined by
\be
B_i\chi_{1\ov m}(z_1,z_2,\dots,z_N)=0\, .
\lab{kzi}
\ee

More generally, in the $n$ component case, the composite bosons will
condense in a collective vacuum defined by
\be
B_i^I\chi_{K}(\{z^J_j\})=0\, ,
\lab{kzI}
\ee
where the new set of operators is given by:
\be
(B_i^I)^{\da} =z_i^I\, , \,\, \,\,\, ~~
B_i^I= \pde_i^I -\pde_i^I\,\, {\sum}^{'} K_{IJ}\,\, \ln(z_i^I-z_j^J) =\pde_i^I
 -{\sum}^{'} {K_{IJ}\over z_i^I-z_j^J}\, .
\lab{FBBI}
\ee \noi
In the case $\nu=1/m$ the vacuum state condition, eq. \ref{kzi}, corresponds to
the Knizhnik-Zamolodchikov (KZ)
linear differential equation for the correlators of an abelian Wess-Zumino
field of conformal weight $m/2$:
\be
\left ( \pde_i -m\sum_{j\neq i}
{1 \over z_i-z_j}\right) \chi_{1\over m}
(z_1, z_2, \dots , z_{N}) =0\, .
\lab{KZi}
\ee \noi
Therefore to each particle we can associate an element of a local
$ \hat {U}(1)$  group of conformal weight $m/2$ that can be written as
a Coulomb gas Vertex Operator (VO)
$V_{\sqrt{m}}(z)=:\exp {[i \sqrt m \phi (z)]}:$\,, where $\phi(z)$
is a ( properly compactified )
free holomorphic scalar field, with the standard mode
expansion
\be
\phi(z)=q-ip\ln z+i\sum_{n\neq 0}\frac{a_n}nz^{-n} .
\ee
As a consequence, the Laughlin gswf can be expressed
as a correlator of VO's:
\begin{equation}
\label{corr}
\chi_{1\over m} (z_1,z_2,\dots,z_N) = \left\langle V_{\sqrt{m}}
\left( z_1\right)\dots V_{\sqrt{m}%
}\left( z_N\right) \right\rangle =
\left\langle N\sqrt m \left | V_{\sqrt{m}}\left( z_1\right) \dots V_{\sqrt{m}%
} \left ( z_N\right) \right | 0 \right\rangle ,
\ee
where the  Fock vacuum state, $\left\langle N{\sqrt m} \right | $,
with momentum $N\sqrt m$, has been  introduced to take into account momentum
conservation.
In the generic abelian case the  equation defining the collective bosonic
vacuum has the form
\be
\lab{KZI}
(\pde_i^I
 -{\sum}^{'} {K_{IJ}\over z_i^I-z_j^J})\chi_K(\{z^I_i\})=0\, .
\ee
The discussion relative to the one component case  is easily generalized
 to the KZ equation for
the $n$ component Hall fluid, provided the $K$ matrix is positive definite.
We will make this assumption as only under this condition
a completely consistent description
on higher genus Riemann surfaces can be achieved$^{17}$. This
amounts to disregard Hall fluids for which the edge currents propagate  in
opposite directions$^{18}$. For such fluids the quantization of Hall
conductance is in general an open problem that, for the case of Jain's
sequences, has been solved$^{19, 16}$ by taking the presence of disorder
explicitly into account.
Under the positivity  condition one can introduce a $n$-component vector of
independent holomorphic fields
$\vec{\phi}(z)$ and a set of
$n$-component vectors $\vec \b_I$ such that
$\vec \b_I \cdot \vec \b_J=K_{IJ}$. Then by defining the VO's
$V_{{\vec \b} _I}(z)=:\exp{[i\vec\b_I\cdot\vec\phi(z)]}:$\, and the currents
$J_a(z)=i\pde\phi_a(z)$ one obtains the following operator product expansion
(OPE)
\be
J_a (z)V_{{\vec \b}_I}(w) \sim { (\vec \b _I)_a \ov z-w}V_{{\vec \b}_I}(w)\, .
\ee

Therefore  one recognizes  eq. \ref{KZI}
as the KZ equation for a $\hat {U}(1)^n$ Wess-Zumino model.
The corresponding correlators will be given by
\be
\lab{CORR}
\chi_{K}(\{z^I_i\})=
\left\langle \prod_{I=1}^n\prod_{i=1}^{N_I}V_{\vec \b
_I}(\{ z_i^I \}) \right\rangle\,.
\ee

We turn now to the more interesting case relative to the fillings of the Jain's
sequences, where out of the KZ equation will emerge
the presence of an extended $\hat {U}(1) \otimes \hat {SU}(n)$ algebra.
The structure of the $K$
matrix for the Jain's sequences is given by $K_{IJ} = {\d}_{IJ}+ 2p$.
Notice that for the Jain's sequences in all components there is
the same number of electrons $N_I$. This derives from the condition
$N_{\Phi}=K_{IJ}N_J$, where $N_{\Phi}$ is the number of units of
external magnetic flux, that guarantees the cancellation, in the average,
between the external magnetic field and the statistical field of the
composite bosons.

The $K$ matrix can be diagonalized by means of an orthogonal
transformation, $K = O^T K_{diag} O$, where
$K_{diag}=\text{{diag}} (1,\dots,1,2np+1)$.
The matrix element  $ O_{IJ}$ with
$I=1,\dots,n-1$ and $J=1,\dots,n$ are strictly related to the matrix elements
of the diagonal generators of the $SU(n)$ group in the fundamental
representation:
\be
\lab{O}
O_{aJ}=t^a_{JJ}\equiv ({\vec u}_J)_a\, ,\quad \quad a=1,\dots,n-1,
\ee
(we have introduced the vectors ${\vec u}_J$
in order to simplify the notation).
The remaining elements are given by $ O_{nJ}=1/\sqrt {n}$, $J=1,\dots,n$.
The explicit form of the matrix $t^a$ is given by
\be
\lab{D}
t^a= {1 \over \sqrt{a(a+1)}}\text{{diag}}(1,\dots,1,-a,0,\dots,0)\, ,
\quad a=1,\dots\, n-1\,.
\ee
Notice that $t^a$ is traceless, as it should.
We can then rewrite $K_{IJ}$ as follows:
\be
K_{IJ}={1\ov \nu}+{\vec u}_I\cdot{\vec u}_J={1 \ov \nu}+t^a_{II}
t^a_{JJ}\, ,
\ee
and  express eq. \ref{KZI} in a form that
makes manifest the presence of an extended symmetry, namely
\be
\lab{KZIU}
(\pde_i^I
 -{\sum}^{'} {\nu^{-1}\over z_i^I-z_j^J}+ {\sum}^{'}{ t^a_{II}
t^a_{JJ}\ov z_i^I-z_j^J } )\chi_K(\{z_l^L\})=0\, .
\ee
where $\nu $ is the filling factor relative to the Jain's states, $\nu=n/
(2np + 1)$.
Notice that the numerator of the term appearing in the first sum
of eq. \ref{KZIU} corresponds strictly to the $\hat {U}(1)$ case discussed
above, see eq. \ref{KZi}. This suggest to perform an orthogonal
transformation on the $\phi$-fields:
\bey
\Phi_a &=&O_{aj} \phi_{j} \, , \,\,\,\,\, a= 1,2,\dots, n-1\, ,
\\
\label{PHI}
\Phi_+ &=&O_{nj}\phi_j ={1\over \sqrt {n}}(\phi_1 + \phi_2 \dots \phi_n)\, .
\eey
We can then rewrite the energy-momentum tensor and the VO's in terms of the
new fields obtaining
\be
\label{T}
T(z)=-{1\ov 2}:\pde {\vec \phi}(z)\! \cdot \!  \pde{\vec \phi}(z): \,
=-{1\ov 2}: \pde\Phi _+ (z)\pde\Phi _+
(z):-{1\ov 2}: \sum_a^{n-1} \pde \Phi _a (z)  \pde\Phi _a (z):\, ,
\ee
and
\be
\lab{VV}
V_{\vec{\b}_I}(z)=\, :\exp{[{\vec \b}_I\cdot {\vec \phi}(z)]}:\, = \,
:e^{ {i \over {\sqrt\nu}}
\Phi_+ (z)}:\,\,:e^{i\vec{u}_I\cdot\vec{\Phi}(z)}:\,
\equiv V_+(z)V_{{\vec u}_I}(z)\, .
\ee
We identify  $i \pde\Phi _+(z)$ as a $\hat {U}(1)$ current and
$J_a (z)= i\pde \Phi _a (z) $ as the diagonal currents of $\hat{SU}(n)$. Indeed
\be
J_a (z)V_{{\vec u}_I}(w) \sim { t^a_{II}\ov z-w}V_{{\vec u}_I}(w)\, .
\ee
This identifies the coefficients $t^a_{II}$ in the second sum of the KZ
eq. \ref{KZIU} as the correct representation matrix elements.
Although the non-diagonal terms do not appear in eq. \ref{KZIU},
as they are absent in the energy-momentum tensor eq. \ref{T},
they are easily introduced as bilinear in the VO's$^{20}$.
As the roots of $SU(n)$ are of the form $\vec \a ={\vec u}_L -{\vec u}_M$,
one has $J_{\vec \a}=:V_{{\vec u}_L}V_{{\vec u}_M}^{\dagger}:$
and their action on the VO
$V_{{\vec u}_I}$ is given by
\be
J_{\vec \a} (z) V_{{\vec u}_I}(w) \sim { \d_{IM} \ov z-w} V_{{\vec u}_L}\, .
\ee
In terms of the new fields the gswf characterized
by the  KZ equation takes the form
\be
\lab{}
\chi_{K}(\{z^I_i\})=
\left\langle 0 \left| \prod_{I=1}^n\prod_{i=1}^{N_I}V_{{\vec u }_I }
( z_i^I )\right | 0 \right\rangle
\left\langle \sum_{I,i} \sqrt {{\nu}^{-1}}\left|\prod_{I=1}^n\prod_{i=1}^{N_I}
V_{+} (z_i^I)\right |0 \right\rangle .
\ee

Notice that the correlator factorizes and, while the $\hat {U}(1)$ factor
carries the full information on the total momentum associated with the VO's,
the $\hat {SU}(n) $ factor automatically satisfies the
 Coulomb gas  neutrality condition.
This implies, as it can be easily seen in terms of the
equivalent plasma picture,
that the $\hat {U}(1)$ mode is charged while the $\hat{SU}(n)$
modes are neutral. The same result can also be obtained by recalling that
the 2DCFT momentum is only rescaled with respect to the physical charge or
by explicitly evaluating the Hall conductance$^{17}$.

The difference between a generic $n$ component Hall fluid and one
relative to the Jain's sequences has important consequences on the structure
of the theory on a non trivial compact Riemann surface and on the nature of
the edge excitations.
It has been shown that while in the generic case there are $n$ sectors of
charged edge excitations, in the Jain's case only the  $\hat {U} (1)$
mode is charged while the remaining $n-1$ are neutral$^{15,16}$.
The same result has been obtained by analyzing the gswf's on a torus, in
ref 17, where the correspondence between gswf's on genus one Riemann
surface and edge states is discussed in the framework of 2DCFT, showing that
both are
characterized by the same integer lattice $\text{{\bf Z}}^n/K \text{{\bf
Z}}^n$. The peculiar nature
of the Jain's sequences is then related to the fact that the relative
lattice can be recast in the much simpler form
$\text{{\bf Z}}/(\det K)\text{{\bf Z}}$, in complete
analogy to the case of a one-component fluid. Whether this characterization
is physically relevant in explaining the observed prominence of Jain's
sequences is an open and interesting problem.

{\bf Acknowledgements}
 A grant by MURST and  the  EEC contract n. SC1-CT92-0789 are acknowledged.

\begin{center}
{\bf References}
\end{center}

\begin{enumerate}

\item  R. B. Laughlin, Phys. Rev. Lett. {\bf 50,} 1395 (1983).

\item  See e.g. F. Wilczek, {\it Fractional Statistics and Anyon
Superconductivity }, World Scientific Singapore (1990).

\item See M. Stone,
{\it Quantum Hall Effect},  World Scientific Singapore (1992), Chap. 3 and
reprints contained therein.

\item  For a review see G. Cristofano, G. Maiella, R. Musto and F. Nicodemi,
Nucl. Phys. {\bf 33C}, 119 (1993).

\item  See e.g. J. Fr\"olich and A. Zee, Nucl. Phys. {\bf B364}, 517 (1991);
X.G. Wen and A. Zee, Phys. Rev. {\bf B 46}, 2290 (1992).

\item  S. H. Simon and B. I. Halperin, Phys. Rev. {\bf B 48} 17368 (1993),
{\bf B 50} 1807 (1994).

\item  J. K. Jain, Phys. Rev. Lett. {\bf  65}, 199 (1989); Phys. Rev. {\bf B
40}, 8079 (1989); {\bf B 41} 7653 (1990).

\item   R. R. Du, H. L. Stormer, D. C. Tsui, L. N. Pfeiffer  and  K. W. West,
Phys. Rev. Lett. {\bf 70}, 2944 (1993).

\item  See e.g. B. I. Halperin, P. A . Lee and N. Read, Phys. Rev. {\bf B
47}, 7312 (1993); R. L. Willett, R. R. Ruel, M. A. Paalanen,  K. W. West and
L. N. Pfeiffer,  Phys. Rev. {\bf B  47}, 7344  (1993); J. K. Wang and V. J.
Goldman, Phys. Rev. Lett. {\bf 67}, 74 (1991)

\item  G. Cristofano, G. Maiella, R. Musto and F. Nicodemi, {\it FQHE and
Jain's approach on the torus}, Int. J. Mod. Phys. {\bf B}, in press.

\item  M. Flohr and R. Varnhagen, Jour. Phys. {\bf A27}, 3999 (1994);
D. Karabali, Nucl. Phys. {\bf FS B419}, 437 (1994).

\item  A. Cappelli, C. A. Trugenberger and G. R. Zemba, Nucl. Phys. {\bf B396}
465 (1993); Phys. Lett. {\bf B 306 }, 100 (1993); Phys. Rev. Lett. {\bf
72} 1902 (1994).

\item S. Iso, D. Karabali and B. Sakita, Phys. Lett. {\bf B296} 143 (1992).

\item  V. G. Knizhnik and A. B. Zamolodchikov, Nucl. Phys. {\bf B247}, 83
(1984).

\item  A. Cappelli, C. A. Trugenberger and G. R. Zemba, {\it Stable
Hierarchical Quantum Hall Fluids as W$_{1+\infty }$ Minimal Models}, preprint
UGVA-DPT 1995/01-879, DFTT 09/95, hep-th 9502021.

\item  C.L. Kane and M. P. A. Fisher, {\it Impurity scattering and transport of
fractional Quantum Hall edge states}, preprint cond-mat 9409028.

\item A. De Martino and R. Musto, {\it Abelian Hall Fluids and Edge States: a
Conformal Field Theory Approach}, preprint DPS-NA-19/95;
INFN-NA-IV-19/95; DFUPG-101/95; cond-mat 9505027.

\item see e.g. X. G. Wen, Int. J. Mod. Phys. {\bf B6} (1992) 1711 and
references contained therein.

\item C. L. Kane, M. P. A. Fisher and J. Polchinski, Phys. Rev. Lett. {\bf 72},
4129 (1994).

\item  P. Goddard and D. Olive, Int. J. Mod. Phys. {\bf A1}, 303 (1986).

\end{enumerate}

\end{document}